\documentclass[letters,fleqn,usenatbib]{mnras}
\usepackage{txfonts}
\usepackage[T1]{fontenc}
\usepackage{ae,aecompl}
\usepackage{graphicx}	
\usepackage{amssymb}	
\usepackage{lscape}
\usepackage{rotating}
\usepackage{url}
\usepackage[british]{babel}

\newcommand{\lppr}{\stackrel{<}{\scriptstyle \sim}}
\newcommand{\lappr}{\raisebox{-0.4ex}{$\lppr$}}
\newcommand{\Mwd}{\mbox{$\mathrm{M_{WD}}$}}
\newcommand{\Msec}{\mbox{$\mathrm{M_{sec}}$}}
\newcommand{\Mtot}{\mbox{$\mathrm{M_{TOT}}$}}
\newcommand{\Msun}{\mbox{$\mathrm{M_{\odot}}$}}
\newcommand{\Rsun}{\mbox{$\mathrm{R_{\odot}}$}}
\newcommand{\Teff}{\mbox{$\mathrm{T_{eff}}$}}

\title[]{The origin of single low-mass WDs: another problem that consequential angular momentum loss in CVs might solve}

\author[Zorotovic \& Schreiber]{M. Zorotovic,$^{1}$\thanks{E-mail: mzorotovic@dfa.uv.cl (MZ)}, M.R. Schreiber,$^{1}$\\
$^{1}$ Instituto de F\'isica y Astronom\'ia, Universidad de Valpara\'iso, Av. Gran Breta\~na 1111, Valpara\'iso, Chile}

\date{Accepted 2016 November 21. Received 2016 November 18; in original form 2016 September 30}

\pubyear{2016}

\begin{document}
\label{firstpage}
\pagerange{\pageref{firstpage}--\pageref{lastpage}}
\maketitle

\begin{abstract}
Low-mass helium-core white-dwarfs (WDs) with masses below $0.5 \Msun$ are
known to be formed in binary star systems but unexpectedly a significant
fraction of them seem to be single. On the other hand, in Cataclysmic
Variables (CVs) a large number of low-mass WD primary stars is predicted but
not observed. We recently showed that the latter problem can be solved if
consequential angular momentum loss causes especially CVs with low-mass WDs to
merge and form single stars. Here we simulate the population of single WDs
resulting from single star evolution and from binary star mergers taking into
account these new merging CVs. We show that according to the revised model of
CV evolution, merging CVs might be the dominant channel leading to the formation of
low-mass single WDs and that the predicted relative numbers are
consistent with observations. This can be interpreted as further evidence for
the revised model of CV evolution we recently suggested. This model includes 
consequential angular momentum loss that increases with decreasing WD mass 
and might not only explain the absence of low-mass WD primaries in CVs but 
also the existence of single low-mass WDs.

 \end{abstract}

\begin{keywords}
white dwarfs -- binaries: close -- novae, cataclysmic variables -- 
\end{keywords}



\section{Introduction}\label{s:intro}

The main sequence (MS) lifetime of the progenitors of helium-core white dwarfs
(WDs) with masses below $0.5 \Msun$, hereafter  
\textit{low-mass WDs}, largely exceeds the Hubble time. However, approximately
10 per cent of all WDs in the solar neighbourhood  
are low-mass WDs \citep[e.g.][]{kepleretal07-1}. The
dominant evolutionary channel producing these  
low-mass WDs is common envelope (CE) evolution of close binary stars, which
can cause the ejection of the envelope of a giant star 
much earlier than it 
is possible in standard single star evolution. 
If the WD progenitor fills its Roche-lobe on the first 
giant branch, dynamically unstable mass transfer leads to the formation of a
gaseous envelope around the helium core of the  
giant and the companion star, which is expelled at the expense of
orbital energy and angular momentum within  
$\lappr\,10^3$yrs \citep[e.g.][]{webbink08-1}.

Indeed, observations show that low-mass WDs are typically members of close
binary stars with either a MS star companion  
\citep{rebassa-mansergasetal11-1} or a second WD in a close orbit
\citep{marshetal95-1}. However, while the binary 
fraction is 100 per cent for extremely low-mass 
WDs ($\Mwd<0.25\Msun$, e.g.
\citealt{brownetal10-1}), there exists a  
population of single low-mass WDs with masses $0.25\Msun<\Mwd<0.50\Msun$
which makes up $\lappr\,20-30$ per cent of all low-mass WDs \citep{brownetal11-1}. 
Combined with the fraction of 
low-mass WDs 
in the solar neighbourhood of $\sim\,10$ per cent 
\citep{kepleretal07-1}, we estimate that 
$\lappr\,2-3$ per cent of all WDs are single low-mass WDs.  
These WDs cannot have formed 
through normal single star evolution as the 
Universe is simply not old enough. 

Several explanations have been put forward to explain the existence of single
low-mass WDs. A rather obvious possibility 
are merger events either during CE evolution 
or of two very low-mass
WDs. However, when  
compared with observed samples these merger channels predict too few low-mass
WDs and far too many single subdwarf B stars  
\citep{nelemans10-1,brownetal11-1}. 

As an alternative option it has been suggested 
almost two decades ago that close-in  
massive planets could lead to the 
formation of single low-mass WDs. In this
scenario the stellar envelope of a $\sim1\Msun$  
star on the first giant branch is expelled in a CE like event but with the 
substellar companion either getting evaporated 
or merging 
with the giant \citep{nelemans+tauris98-1}. While this is in general possible,
rather massive planets or brown dwarfs are  
needed to unbind a substantial fraction of the envelope of the giant and it is
not clear if massive close-in companions  
exist in large enough numbers to explain the fraction of low-mass WDs that
seem to be single. 

As a third possibility strong mass 
loss or spontaneous ejection 
of the envelope close to the tip of
the first giant branch for metal-rich stars 
has been proposed
as an explanation for the existence of 
single low-mass WDs \citep{kilicetal07-1,
brownetal11-1,hanetal94-1,mengetal08-1}. 
This scenario is consistent with 
the colour-magnitude 
diagram for the metal-rich cluster NGC\,6791, which 
can be explained by a WD population dominated 
by low-mass WDs \citep{hansen05-1,kaliraietal07-1}. 
However, a convincing alternative explanation for the 
colour-magnitude distribution of the WDs in this 
cluster was developed by 
\citet{althausetal10-1} and \citet{garcia-berroetal11-1}
who incorporated e.g. the energy released by the 
processes of $^{22}$Ne sedimentation and 
constrained the number of single helium-core WDs in 
NGC\,6791 to below 5 per cent. 
In addition, \citet{miglioetal12-1} measured mass 
loss on the giant branch in NGC\,6791 and found values 
much too small to produce single low-mass WDs.

Finally, some single low-mass WDs might 
be the remnants of the companion stars to 
WDs that exploded as type Ia Supernovae.  
These supernovae evolved through the 
single-degenerate channel, either with red giant 
donors \citep{justhametal09-1} whose envelopes have been 
stripped off by the explosion,
or with close helium-star companions that might become 
low-mass WDs \citep{wang+han09-1}. 
Indeed, recent observations of the hypervelocity 
helium-star US\,708, with a mass of 0.3\Msun, suggest that this is 
the most likely scenario for its 
formation \citep{geieretal15-1}. However, as it is not 
even clear to what extent type Ia Supernovae are produced 
by the single-degenerate channel, it remains
elusive that we will soon be able to reliably 
estimate the formation rates
of single low-mass WDs through these 
channels.

We here suggest a completely new formation mechanism 
for single low-mass
WDs which is based on the new model for the evolution 
of cataclysmic variables (CVs) proposed by \citet{schreiberetal16-1}. CVs are
semi-detached close compact binaries in  
which a WD accretes hydrogen-rich material from its low-mass MS companion star
\citep[e.g.][]{kniggeetal11-1}. Since the  
first binary population models for CVs have been published, a large number of
CVs with low-mass WD primaries is predicted  
\citep[e.g.][]{dekool92-1,hanetal95-1,politano96-1} but not a single
convincing candidate has been found observationally, which cannot be explained by 
selection effects \citep{zorotovicetal11-1,wijnenetal15-1}. 
We recently showed that the problem
of the missing low-mass WDs in CVs can be solved  
if additional angular momentum loss generated by the mass transfer itself,
so-called consequential angular momentum loss (CAML),  
causes especially CVs with low-mass WDs to merge
\citep{schreiberetal16-1}. This empirical approach
represents an attractive possibility as other  
discrepancies between CV evolutionary models and observational population studies can be solved simultaneously 
\citep[see also][]{nelemansetal16-1}. Amazingly, the new model may also solve
a problem that is at first glance not related 
to CV evolution: The predicted large numbers of merging  
CVs should produce significant numbers of single low-mass WDs which might
explain their existence. In other words, the CVs with low-mass WDs we seem to be unable 
to find might have merged and evolved into the single low-mass WDs we have no
convincing explanation for. In this letter we present 
binary population models that test this  
hypothesis and find that indeed merging
CVs might be the dominant channel producing  
single low-mass WDs.  

\section{Binary population models}

In order to predict the mass distribution of single WDs including those that
formed through merging CVs, we performed population
models of both single and binary stars. Assuming an 
initial binary fraction of 50 per cent, we generated an initial population of $10^7$ 
single stars and the same number of MS+MS binaries. We used the initial-mass function of
\citet{kroupaetal93-1} in the range $0.8-10\Msun$ 
for the mass distributions of single stars and primary stars in binaries, and a
constant star formation rate with an upper limit  
of 13.5 Gyrs for the age of the Galaxy \citep{pasquinietal04-1}. 
In the case of binaries, we assumed a flat initial mass-ratio distribution 
for secondary masses \citep{sanaetal09-1}, with a lower limit of 
$\Msec=0.069\Msun$ (the brown dwarf limit in \citealt{kniggeetal11-1}). We 
also used a flat distribution in $\log a$  
ranging from 3 to $10^6\Rsun$ for the orbital separations
\citep{popovaetal82-1}, and no eccentricity. 
Solar metallicity was assumed for all stars.

The single-star evolution (SSE) and the binary-star evolution (BSE) codes from \citet{hurleyetal00-1,hurleyetal02-1} were 
used to evolve single and binary stars, respectively. The CE phase was modelled with an efficiency of $\alpha_\mathrm{CE} = 0.25$ 
\citep{zorotovicetal10-1} and with the binding energy parameter $\lambda$ properly computed assuming no contributions of 
recombination energy \citep[see Section\,2.2 in][]{zorotovicetal14-2}. When a close WD+MS binary was formed, its subsequent 
evolution was calculated with our own code \citep{zorotovicetal16-1}. All zero-age WD+MS binaries 
were evolved to their current age assuming disrupted magnetic braking.
CV evolution was calculated taking into account empirical CAML as suggested by 
\citet{schreiberetal16-1}. This revised model predicts that in a large number of
systems with low-mass WDs unstable mass transfer is generated leading to a
merger event instead of a CV with stable mass transfer. 

\begin{figure*}
\includegraphics[width=0.35\textwidth,angle=270]{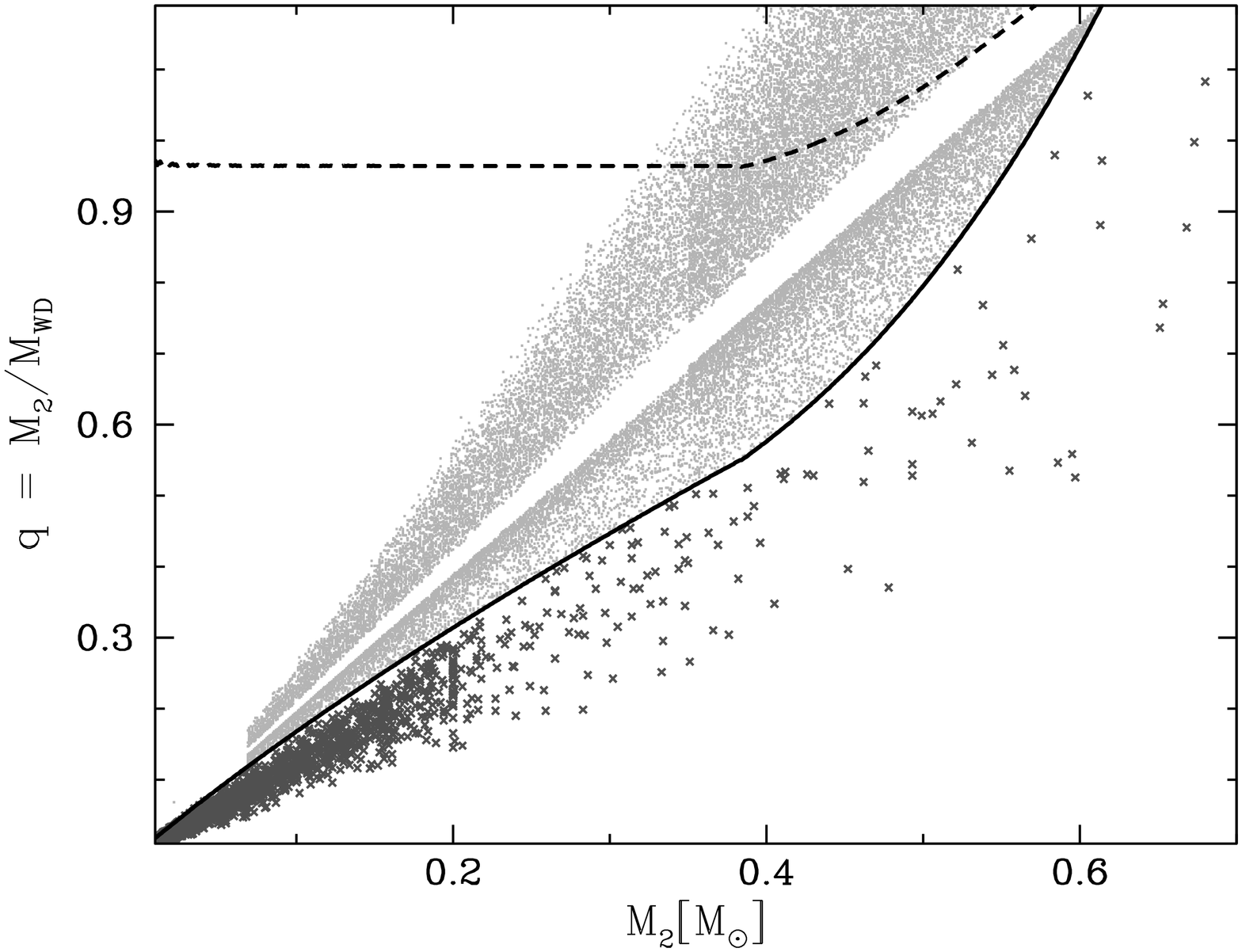}
\includegraphics[width=0.35\textwidth,angle=270]{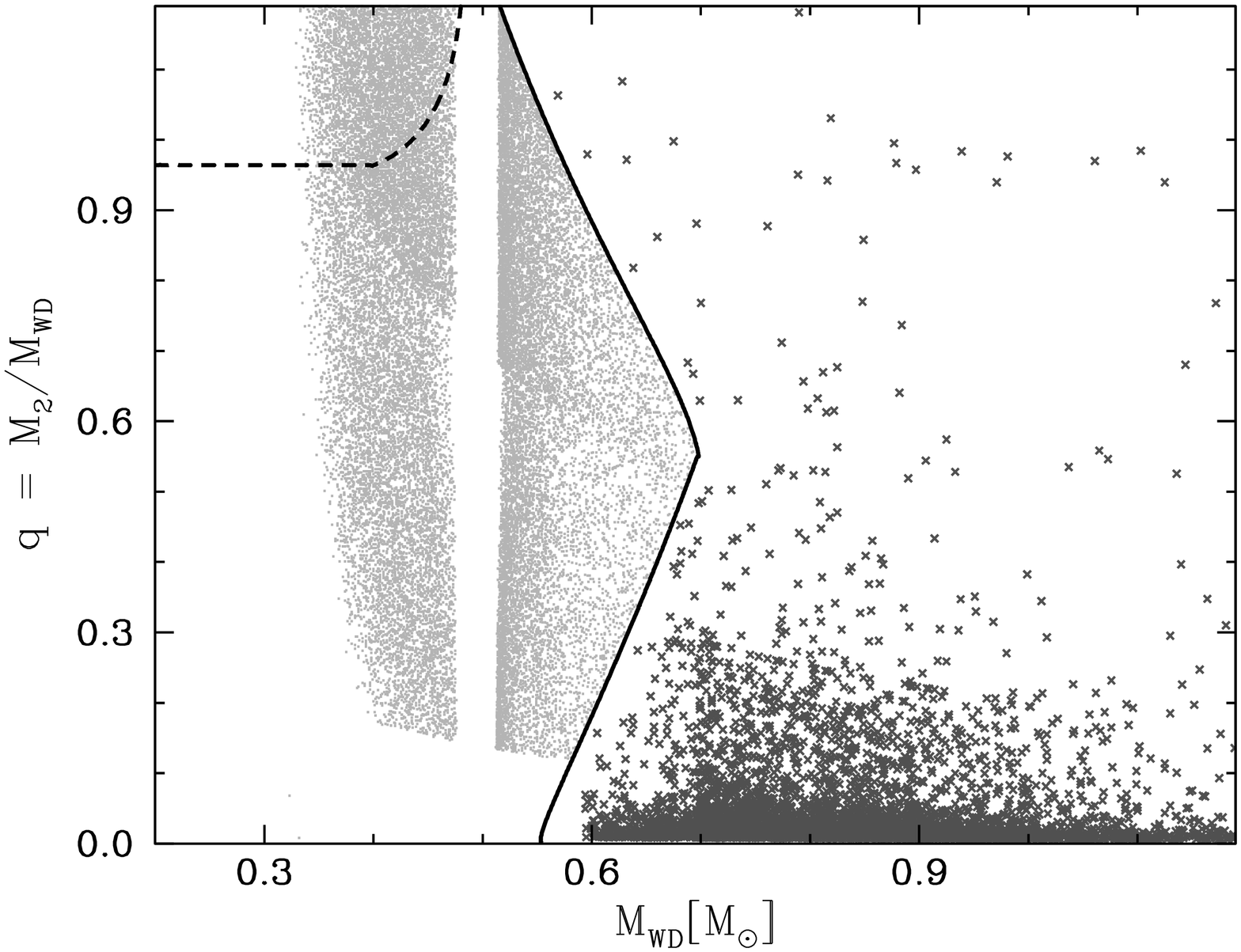}
\caption{Mass-ratio versus secondary star mass (left) and  WD mass (right) for CVs (black) and systems that merge (grey) 
due to unstable mass transfer. The solid line represents the critical mass-ratio for dynamically unstable mass transfer assuming 
empirical CAML as suggested by \citet{schreiberetal16-1} while the dashed line is the critical mass-ratio obtained for classical 
CAML \citep{king+kolb95-1}. The new model predicts significantly more CV mergers than the classical model. Especially systems with 
low-mass WDs merge instead of becoming CVs.}
\label{fig:Mq}
\end{figure*}

\begin{table*}
\caption{Number of single WDs produced by the different channels. ``Others mergers'' include mergers of MS+MS binaries and double 
WDs. The fraction of low-mass WDs with respect to the total number of single
WDs is $\sim1.6$ per cent and $\sim0.6$ per cent for model 
\textit{a} and \textit{b} respectively.}
  \begin{tabular}{ | l | rl | r | r | }
\hline
      & \multicolumn{2}{|c|}{All single WDs} & \multicolumn{2}{|c|}{$\Mwd\leq0.5\Msun$} \\
      channel & Number & Fraction & N. Case \textit{a} & N. Case \textit{b} \\
\hline
\hline
      Single Stars & 4\,698\,675 & 71.4\% & 0       & 0 \\
      CE merger    & 1\,413\,702 & 21.5\% & 7\,544  & 7\,544 \\
      CV mergers   & 143\,532    & 2.2\%  & 93\,636 & 29\,554  \\
      Other mergers & 325\,711    & 4.9\%  & 1\,635  & 1\,635  \\
\hline
      Total & 6\,581\,620 &  & 102\,815 (1.6\%) & 38\,733 (0.6\%) \\
\hline
\label{tab:wds}
  \end{tabular}
\end{table*}

The merging CVs should form single stars that then evolve into single WDs but this process 
is very complex. It is clear that following the merger event the systems enter 
a red giant configuration due to a rejuvenation 
of the hydrogen-burning shell, i.e. the disrupted 
secondary forms a giant envelope around 
the WD \citep{shenetal09-1}. However, 
it is not clear if all the mass of the secondary is 
retained as an envelope or if, more likely, a fraction 
of this mass is expelled in a CE-like event. For the 
sake of simplicity, we decided to estimate the 
fraction of low-mass WDs produced by this 
channel considering two cases: \textit{a}) the entire 
mass of the secondary star at the moment of merger is 
expelled, i.e. the mass of the WD remains unchanged; \textit{b}) the entire mass of the secondary at the moment of merger is accumulated 
around the WD as a giant envelope. In case \textit{b} a giant with a total mass of $\Mtot=\Mwd+\Msec$ and a core mass 
of \Mwd\, is generated. If \Mwd\, is larger than the final mass of a star with initial mass \Mtot, we assumed that 
it cannot grow more and therefore the final mass of the WD remains unchanged. If, on the other hand, the mass of the WD is smaller, 
we assumed that it can grow up to a mass given by the final (WD) mass of a single star with initial mass \Mtot. 
This is a 
simplified approach as the WD growth 
depends on the core-mass growth rate and the lifetime 
of the star on the giant branch, 
which are not necessarily equal to those of a star 
with initial mass \Mtot. However, our simple approach 
is fully sufficient to get a reasonable estimate 
of the relative number of low-mass WDs formed by CV 
mergers. Detailed modelling of the merger 
process or the subsequent evolution of the 
resulting giant star are far beyond the scope of this 
paper.

In order to estimate the relative importance of CV 
mergers for the formation
of single low-mass WDs, 
mergers of systems other 
than WD+MS that can produce single WDs were considered too. Merger events during the CE phase or in other close systems like MS+MS 
or double WD binaries were computed with BSE (see section 2.7 in \citealt{hurleyetal02-1} for details). MS+MS mergers contribute
to the population of high-mass WDs, while mergers of double WDs or during the CE phase can form both low-mass and high-mass WDs. 

\section{The predicted single WD mass distribution}

The new model for CV evolution including empirical CAML predicts that in most WD+MS binaries containing low-mass WDs dynamically unstable 
mass transfer is triggered after the MS star fills its Roche-lobe \citep{schreiberetal16-1}. Therefore, significantly more mergers are 
predicted than by the previously assumed much weaker forms of CAML \citep[e.g.][]{king+kolb95-1}. This is illustrated in 
Figure\,\ref{fig:Mq} which shows the mass ratio $q=\Msec/\Mwd$ as a function of secondary star mass (left panel) and WD mass (right panel). 
The limits separating stable and dynamically unstable mass transfer are very different for the new empirical and the classical form of CAML 
(solid and dashed line in Figure\,\ref{fig:Mq}, respectively). In the new model strong CAML for systems with low-mass WDs generates unstable 
mass transfer and the two stars merge. 

\begin{figure}
\includegraphics[width=0.35\textwidth,angle=270]{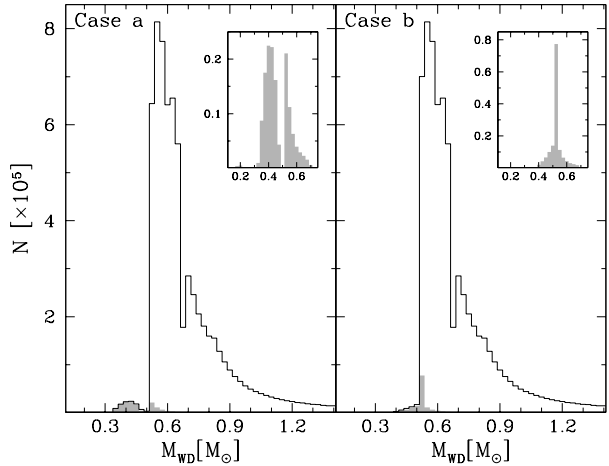}
\caption{WD mass distribution of the single WDs predicted by our simulations. Left and right panels are for case \textit{a} and \textit{b} 
respectively (see text). Single WDs produced by CV mergers are shown in grey
(a zoom is shown in the inset of each panel). 
The vast majority of single low-mass WDs ($\Mwd\leq0.5\Msun$) are formed by CV mergers.}
\label{fig:Mwd}
\end{figure}

The merging CVs generated by the new model enter a giant-like configuration and will become single WDs after the envelope is expelled. 
The exact number of low-mass WDs produced due to CV mergers depends on how much mass is lost during the merging process. We tested two 
prescriptions, i.e. we either assumed that all the mass of the secondary at the moment of merger is expelled (model \textit{a}) 
or that all the mass is accumulated as an envelope around the WD (model \textit{b}). The resulting mass distributions of single WDs for 
both models are shown in Figure\,\ref{fig:Mwd}. These distributions are of course dominated by single WDs descending from initially single 
stars, but significant numbers of low-mass WDs (with $\Mwd\leq0.5\Msun$) are produced by CV mergers (represented by the grey histograms 
in Fig.\,\ref{fig:Mwd}, see also Table\,\ref{tab:wds}). Even if we assume that the whole mass of the secondary at the moment of merger is 
accumulated around the WD, the dominant channel producing single low-mass WDs in our simulation are merging CVs (Table\,\ref{tab:wds}). 

Our simulations are based on several reasonable but uncertain assumptions that could in principal impact the predicted relative number of 
single low-mass WDs produced by CV mergers. Despite some rather convincing 
constraints 
\citep{sanaetal09-1,zorotovicetal10-1}, the most 
important uncertain parameters are probably the initial mass-ratio
distribution and the CE efficiency. However, 
in \citet{zorotovicetal14-1} 
we determined the fraction of low-mass (helium-core) WDs among post common envelop binaries (PCEBs) for a wide range of CE efficiencies and 
two initial mass-ratio distributions and found only very weak dependencies. The fraction of low-mass WDs in PCEBs remained nearly unchanged 
in the range of 40 -- 60 per cent. This implies that also the fraction of low-mass WDs among CVs will not dramatically change if other 
assumptions for the mass-ratio distribution or the CE efficiency are used. Another uncertain parameter is the age of the Galaxy. However, 
assuming 10 Gyrs instead of 13.5 we found negligible changes in the resulting fractions of single low-mass WDs. Finally, the initial binary 
fraction of 50 per cent we assumed for all types of binaries independent of
the primary star mass is probably incorrect. There is growing evidence that 
the binary fraction increases with increasing primary mass
\citep[e.g.][]{lada06-1}. We thus may underestimate the number of descendants 
from more massive binaries which implies that we perhaps slightly
underestimate the number of massive double WDs and maybe also the CE  
mergers of massive stars. Both these channels can produce massive single WDs
and we therefore might slightly overestimate the fraction of  
single low-mass WDs. However, the vast majority of binaries that form single
WDs descend from systems with initial primary star  
masses in the range of $1-2\Msun$, where the binary fraction is supposed to be
similar to the $\sim50$ per cent we assumed  
\citep[e.g.][]{duquennoy91-1}. 
Therefore, our simplified assumption does not
affect the conclusions of this paper.  

We conclude that if the empirical CAML model from \citet{schreiberetal16-1} is
included, the dominant channel producing single low-mass WDs  
might potentially be CV mergers. Our model predicts a fraction of low-mass WDs among single WDs
of the order of one per cent. These findings are independent of model
uncertainties.

\section{The impact of selection effects and observational biases}

The fraction of single low-mass WDs predicted by the new empirical CAML model for CV evolution is of the order of one per cent. However, the 
WD mass distribution (shown in Fig.\,\ref{fig:Mwd}) represents the {\em{total}} population of single WDs predicted by our model and does not 
take into account observational biases and selection effects which need to be considered before our result can be compared with the observations. 

\begin{figure}
\includegraphics[width=0.35\textwidth,angle=270]{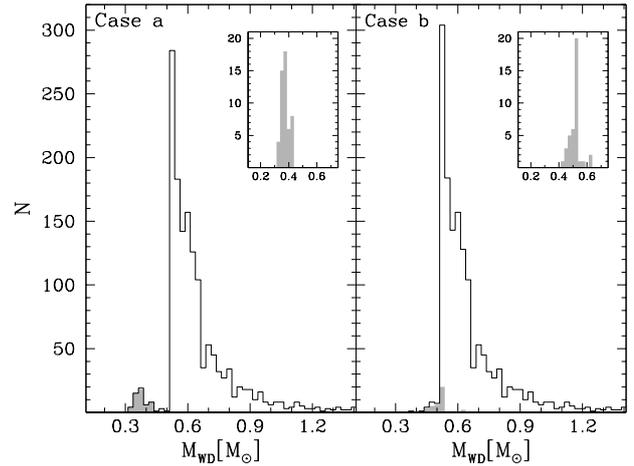}
\caption{Same as in Figure\,\ref{fig:Mwd} but after applying the filters in effective temperature and \textit{g} magnitude from 
\citet{tremblayetal16-1} described in the text.}
\label{fig:MwdTremblay}
\end{figure}

Observational samples such as those from the Sloan Sloan Digital Sky Survey \citep[SDSS, ][]{abazajianetal09-1} are magnitude-limited. To 
construct a magnitude-limited sample from our predicted sample of all single WDs we assigned a distance to each object in our final single WD 
sample using the same galactic distribution as \citet{tremblayetal16-1}. According to these authors the spectroscopic completeness of DA WDs 
from SDSS data release 7 is constant over the parameter ranges $16\,000 < \Teff(K)< 22\,000$ and $16.0< \textit{g} < 18.5$. In order to construct 
a sub-sample of WDs with these parameters we computed the effective temperature of the simulated WDs using the cooling tracks by 
\citet{althaus+benvenuto97-1} for helium-core WDs (if $\Mwd\leq0.5\Msun$) and \citet{fontaineetal01-1} for 
carbon/oxygen-core WDs (if $\Mwd>0.5\Msun$) assuming only thick hydrogen envelopes (DA WDs). Using these temperatures and the WD masses, 
we calculated the absolute magnitude $M_\textit{g}$ using the colour tables from \citet[][ their pure hydrogen grids]{holberg+bergeron06-1}. 
The resulting WD mass distributions are shown in Figure\,\ref{fig:MwdTremblay}. The fraction of single low-mass WDs is slightly increased 
(3.9\% in model \textit{a} and 1.3\% in model \textit{b}) and CV mergers remain the dominant channel producing low-mass WDs, with relative 
contributions nearly identical to the ones obtained before. These results do not change if we somewhat relax the rather strict selection 
criteria. For example, if we take into account WDs with reliable masses from spectral fitting, i.e. WDs with $\Teff > 12\,000$K \citep{kepleretal07-1}, 
and \textit{i} magnitudes covered by the SDSS Quasar 
survey \citep{richardsetal02-1}, i.e. $15.5 < \textit{i} < 19.1$, the fraction of single 
WDs with $\Mwd\leq0.5\Msun$ is 3.3 per cent in model \textit{a} and 1.4
per cent in model \textit{b}. 

The slight increase of the fraction of low-mass WDs if observational selection effects and biases are taken into account is mainly caused by the 
longer MS lifetimes of the progenitors of low-mass WDs, which are relatively low-mass MS stars 
\citep[$M_\mathrm{i}\,\lappr\,1.8\Msun$, see Figure\,2 in][]{zorotovic+schreiber13-1}. This implies that the evolutionary time-scale to form a 
CV with a low-mass WD is on average longer than to form a CV with a more massive WD. Therefore, the fraction of WDs too cool to be detected 
(e.g. by SDSS) is smaller for low-mass WDs because these are in general
younger (and hotter) than more massive WDs. This effect is complemented  
by the fact that high-mass WDs cool faster than low-mass WDs. The final
prediction of our model is thus that in observed samples $\sim1-4$ per cent of all single WDs 
should be low-mass WDs mostly produced by CV mergers.

\section{Conclusion}

Despite some promising suggestions, the existence of single low-mass WDs has
so far not been convincingly explained. In this paper we show that 
if consequential angular momentum loss as described by \citet{schreiberetal16-1} is included in CV evolution, a significant number of single 
low-mass WDs is predicted to result from merging CVs. These single low-mass WDs should make up about 1 -- 4 per cent of all single WDs. 

The only estimate of the fraction of single low-mass WDs derived from
observations are the $\lappr\,20-30$ per cent of single WDs among low-mass 
WDs from \citet{brownetal11-1}. If combined with the finding from \citet{kepleretal07-1} that low-mass WDs make 
up 10 per cent of all WDs in the solar neighbourhood, we find good
agreement with the model predictions. The revised model for CV evolution  
suggested by \citet{schreiberetal16-1} might therefore not only solve several long standing problems of CV evolution. By postulating that especially 
CVs with low-mass WDs merge, this model also offers an explanation for the existence of single low-mass WDs. It might well be that the low-mass WDs 
in CVs we expected to exist but never found, were hidden from us because they evolved into the single low-mass WDs whose existence we always 
struggled to explain.  

\section*{Acknowledgements}
We acknowledge financial support from FONDECYT 
(grants 3130559 and 1141269).


\begin{thebibliography}{48}
\expandafter\ifx\csname natexlab\endcsname\relax\def\natexlab#1{#1}\fi

\bibitem[{{Abazajian} et~al.(2009)}]{abazajianetal09-1}
{Abazajian}, K.~N., et~al., 2009, \apjs, 182, 543

\bibitem[{{Althaus} \& {Benvenuto}(1997)}]{althaus+benvenuto97-1}
{Althaus}, L.~G., {Benvenuto}, O.~G., 1997, \apj, 477, 313

\bibitem[{{Althaus} et~al.(2010)}]{althausetal10-1}
{Althaus}, L.~G., et~al., 2010, \apj, 719, 612

\bibitem[{{Brown} et~al.(2011)}]{brownetal11-1}
{Brown}, J.~M., et~al., 2011, \apj, 730, 67

\bibitem[{{Brown} et~al.(2010)}]{brownetal10-1}
{Brown}, W.~R., et~al., 2010, \apj, 723, 1072

\bibitem[{{de Kool}(1992)}]{dekool92-1}
{de Kool}, M., 1992, \aap, 261, 188

\bibitem[{{Duquennoy} \& {Mayor}(1991)}]{duquennoy91-1}
{Duquennoy}, A., {Mayor}, M., 1991, \aap, 248, 485

\bibitem[{{Fontaine}, {Brassard} \& {Bergeron}(2001)}]{fontaineetal01-1}
{Fontaine}, G., {Brassard}, P., {Bergeron}, P., 2001, PASP, 113, 409

\bibitem[{{Garc{\'{\i}}a-Berro} et~al.(2011)}]{garcia-berroetal11-1}
{Garc{\'{\i}}a-Berro}, et~al., 2011, \aap, 533, A31

\bibitem[{{Geier} et~al.(2015)}]{geieretal15-1}
{Geier}, S., et~al., 2015, Science, 347, 1126

\bibitem[{{Han}, {Podsiadlowski} \& {Eggleton}(1994)}]{hanetal94-1}
{Han}, Z., {Podsiadlowski}, P., {Eggleton}, P.~P., 1994, \mnras, 270, 121

\bibitem[{{Han}, {Podsiadlowski} \& {Eggleton}(1995)}]{hanetal95-1}
{Han}, Z., {Podsiadlowski}, P., {Eggleton}, P.~P., 1995, \mnras, 272, 800

\bibitem[{{Hansen}(2005)}]{hansen05-1}
{Hansen}, B.~M.~S., 2005, \apj, 635, 522

\bibitem[{{Holberg} \& {Bergeron}(2006)}]{holberg+bergeron06-1}
{Holberg}, J.~B., {Bergeron}, P., 2006, \aj, 132, 1221

\bibitem[{{Hurley}, {Pols} \& {Tout}(2000)}]{hurleyetal00-1}
{Hurley}, J.~R., {Pols}, O.~R., {Tout}, C.~A., 2000, \mnras, 315, 543

\bibitem[{{Hurley}, {Tout} \& {Pols}(2002)}]{hurleyetal02-1}
{Hurley}, J.~R., {Tout}, C.~A., {Pols}, O.~R., 2002, \mnras, 329, 897

\bibitem[{{Justham} et~al.(2009)}]{justhametal09-1}
{Justham}, S., et~al., 2009, \aap, 493, 1081

\bibitem[{{Kalirai} et~al.(2007)}]{kaliraietal07-1}
{Kalirai}, J.~S., et~al., 2007, \apj, 671, 748

\bibitem[{{Kepler} et~al.(2007)}]{kepleretal07-1}
{Kepler}, S.~O., et~al., 2007, \mnras, 375, 1315

\bibitem[{{Kilic}, {Stanek} \& {Pinsonneault}(2007)}]{kilicetal07-1}
{Kilic}, M., {Stanek}, K.~Z., {Pinsonneault}, M.~H., 2007, \apj, 671, 761

\bibitem[{{King} \& {Kolb}(1995)}]{king+kolb95-1}
{King}, A.~R., {Kolb}, U., 1995, \apj, 439, 330

\bibitem[{{Knigge}, {Baraffe} \& {Patterson}(2011)}]{kniggeetal11-1}
{Knigge}, C., {Baraffe}, I., {Patterson}, J., 2011, \apjs, 194, 28

\bibitem[{{Kroupa} ,{Tout} \& {Gilmore}(1993)}]{kroupaetal93-1}
{Kroupa}, P., {Tout}, C.~A., {Gilmore}, G., 1993, \mnras, 262, 545

\bibitem[{{Lada}(2006)}]{lada06-1}
{Lada}, C.~J., 2006, \apjl, 640, L63

\bibitem[{{Marsh}, {Dhillon} \& {Duck}(1995)}]{marshetal95-1}
{Marsh}, T.~R., {Dhillon}, V.~S., {Duck}, S.~R., 1995, \mnras, 275, 828

\bibitem[{{Meng}, {Chen} \& {Han}(2008)}]{mengetal08-1}
{Meng}, X., {Chen}, X., {Han}, Z., 2008, \aap, 487, 625

\bibitem[{{Miglio} et~al.(2012)}]{miglioetal12-1}
{Miglio}, A., et~al., 2012, \mnras, 419, 2077

\bibitem[{{Nelemans}(2010)}]{nelemans10-1}
{Nelemans}, G., 2010, \apss, 329, 25

\bibitem[{{Nelemans} \& {Tauris}(1998)}]{nelemans+tauris98-1}
{Nelemans}, G., {Tauris}, T.~M., 1998, \aap, 335, L85

\bibitem[{{Nelemans} et~al.(2016)}]{nelemansetal16-1}
{Nelemans}, G., et~al., 2016, \apj, 817, 69

\bibitem[{{Pasquini} et~al.(2004)}]{pasquinietal04-1}
{Pasquini}, L., et~al., 2004, \aap, 426, 651

\bibitem[{{Politano}(1996)}]{politano96-1}
{Politano}, M., 1996, \apj, 465, 338

\bibitem[{{Popova}, {Tutukov} \& {Yungelson}(1982)}]{popovaetal82-1}
{Popova}, E.~I., {Tutukov}, A.~V., {Yungelson}, L.~R., 1982, \apss, 88, 55

\bibitem[{{Rebassa-Mansergas} et~al.(2011)}]{rebassa-mansergasetal11-1}
{Rebassa-Mansergas}, A., et~al., 2011, \mnras, 413, 1121

\bibitem[{{Richards} et~al.(2002)}]{richardsetal02-1}
{Richards}, G.~T., et~al., 2002, \aj, 123, 2945

\bibitem[{{Sana}, {Gosset} \& {Evans}(2009)}]{sanaetal09-1}
{Sana}, H., {Gosset}, E., {Evans}, C.~J., 2009, \mnras, 400, 1479

\bibitem[{{Schreiber}, {Zorotovic} \& {Wijnen}(2016)}]{schreiberetal16-1}
{Schreiber}, M.~R., {Zorotovic}, M., {Wijnen}, T.~P.~G., 2016, \mnras, 455, L16

\bibitem[{{Shen}, {Idan} \& {Bildsten}(2009)}]{shenetal09-1}
{Shen}, K.~J., {Idan}, I., {Bildsten}, L., 2009, \apj, 705, 693

\bibitem[{{Tremblay} et~al.(2016)}]{tremblayetal16-1}
{Tremblay}, P.-E., et~al., 2016, \mnras, 461, 2100

\bibitem[{{Wang} \& {Han}(2009)}]{wang+han09-1}
{Wang}, B., {Han}, Z., 2009, \aap, 508, L27

\bibitem[{{Webbink}(2008)}]{webbink08-1}
{Webbink}, R.~F., 2008, in {Milone~E.~F., Leahy~D.~A., \& Hobill~D.~W.}, eds.,
  Astrophysics and Space Science Library, Vol. 352, Springer, Berlin, p. 233
  
\bibitem[{{Wijnen}, {Zorotovic} \& {Schreiber}(2015)}]{wijnenetal15-1}
{Wijnen}, T.~P.~G., {Zorotovic}, M., {Schreiber}, M.~R., 2015, \aap, 577, A143
  
\bibitem[{{Zorotovic} \& {Schreiber}(2013)}]{zorotovic+schreiber13-1}
{Zorotovic}, M., {Schreiber}, M.~R., 2013, \aap, 549, A95

\bibitem[{{Zorotovic}, {Schreiber} \& {G{\"a}nsicke}(2011)}]{zorotovicetal11-1}
{Zorotovic}, M., {Schreiber}, M.~R., {G{\"a}nsicke}, B.~T., 2011, \aap, 536, A42

\bibitem[{{Zorotovic}, {Schreiber} \& {Parsons}(2014)}]{zorotovicetal14-2}
{Zorotovic}, M., {Schreiber}, M.~R., {Parsons}, S.~G., 2014, \aap, 568, L9

\bibitem[{{Zorotovic} et~al.(2010)}]{zorotovicetal10-1}
{Zorotovic}, M., et~al., 2010, \aap, 520, A86

\bibitem[{{Zorotovic} et~al.(2014)}]{zorotovicetal14-1}
{Zorotovic}, M., et~al., 2014, \aap, 568, A68

\bibitem[{{Zorotovic} et~al.(2016)}]{zorotovicetal16-1}
{Zorotovic}, M., et~al., 2016, \mnras, 457, 3867

\end{thebibliography}
%

\bsp	
\label{lastpage}
\end{document}